\documentclass{iaus}
\usepackage{graphicx}

\title[GHOSTS: The Resolved Stellar Outskirts of Massive Disk Galaxies] %% give here short title %%
{GHOSTS: The Resolved Stellar Outskirts of Massive Disk Galaxies}

\author[Roelof S.\ de Jong, Anil Seth, and the GHOSTS team]   %% give here short author list %%
{Roelof S.\ de Jong$^1$, A.C.\ Seth$^2$, E.F.\ Bell$^3$, T.M.\
  Brown$^1$,\\ J.S.\ Bullock$^4$, S.\ Courteau$^5$, J.J.\ Dalcanton$^6$,
H.C.\ Ferguson$^1$,\\ P.\ Goudfrooij$^1$, S.\ Holfeltz$^1$, C.\ Purcell$^4$,\\
D.\ Radburn-Smith$^1$, D.\ Zucker$^7$}

\affiliation{
$^1$Space Telescope Science Institute, 3700 San Martin Drive, Baltimore, MD 21218, USA\\
$^2$CFA, 
$^3$MPIA, 
$^4$UC Irvine, 
$^5$Queen's Univ., 
$^6$Univ. of Washington,
$^7$Univ.\ of Cambridge%
}

\pubyear{2007}
\volume{241}  %% insert here IAU Symposium No.
\pagerange{119--126}
\date{?? and in revised form ??}
\jname{Stellar Populations as Building Blocks of Galaxies}
\editors{A.\ Vazdekis and R.\ Peletier, eds.}
\begin{document}

\maketitle

\begin{abstract}
  We show initial results from our ongoing HST/ACS
  GHOSTS survey of the resolved stellar
  envelopes of 14 nearby, massive disk galaxies. In hierarchical
  galaxy formation the stellar halos and thick disks of galaxies are
  formed by accretion of minor satellites and therefore contain
  valuable information about the (early) assembly process of
  galaxies. We detect for the first time the very small halo of
  NGC\,4244, a low mass edge-on galaxy. We find that
  massive galaxies have very extended halos, with equivalent surface
  brightnesses of 28-29 $V$-mag arcsec$^{-2}$ at 20-30\,kpc from the
  disk. The old RGB stars of the thick disk in the NGC\,891 and
  NGC\,4244 edge-on galaxies truncate at the same radius as the young
  thin disk stars, providing insights into the formation of both disk
  truncations and thick disks.  We furthermore present the stellar
  populations of a very low surface brightness stream around M83, the
  first such a stream resolved into stars beyond those of the Milky
  Way and M31.

\end{abstract}

%\section{Introduction}

In recent years we have started to appreciate that the outer banks of
galaxies contain valuable information about the formation process of
galaxies. In hierarchical galaxy formation the stellar halos and thick
disks of galaxies are formed by accretion of minor satellites,
predominantly in the earlier assembly phases. The size, metallicity,
and amount of substructure in current day halos are therefore directly
related to issues like the small scale properties of the primordial
power spectrum of density fluctuations and the suppression of star
formation in small dark matter halos after reionization.

To exploit this information we have started the GHOSTS\footnote{GHOSTS: Galaxy Halos, Outer disks, Substructure,
    Thick disks and Star clusters\hfill} survey, which
will sample the resolved stellar populations along the major and minor
axes of 14 nearby galaxies using HST/ACS and WFPC2. Our data provide
color-magnitude diagrams 1.5-2.5 magnitudes below the tip of the Red
Giant Branch.  We measure the stellar density distribution from star
counts down to very low average surface brightnesses, equivalent to
$\sim$32 V-mag per square arcsec.  We will also obtain spatial
information on the metallicity distributions of the Red Giant Branch
stars. Our targets have large angular extents and we need several
images to sample one principal axis. For the galaxies where we
received enough data to create radial profiles, the results have been
both remarkable and highly varied.

\begin{figure}
\begin{minipage}{0.47\linewidth}
\includegraphics[width=\linewidth]{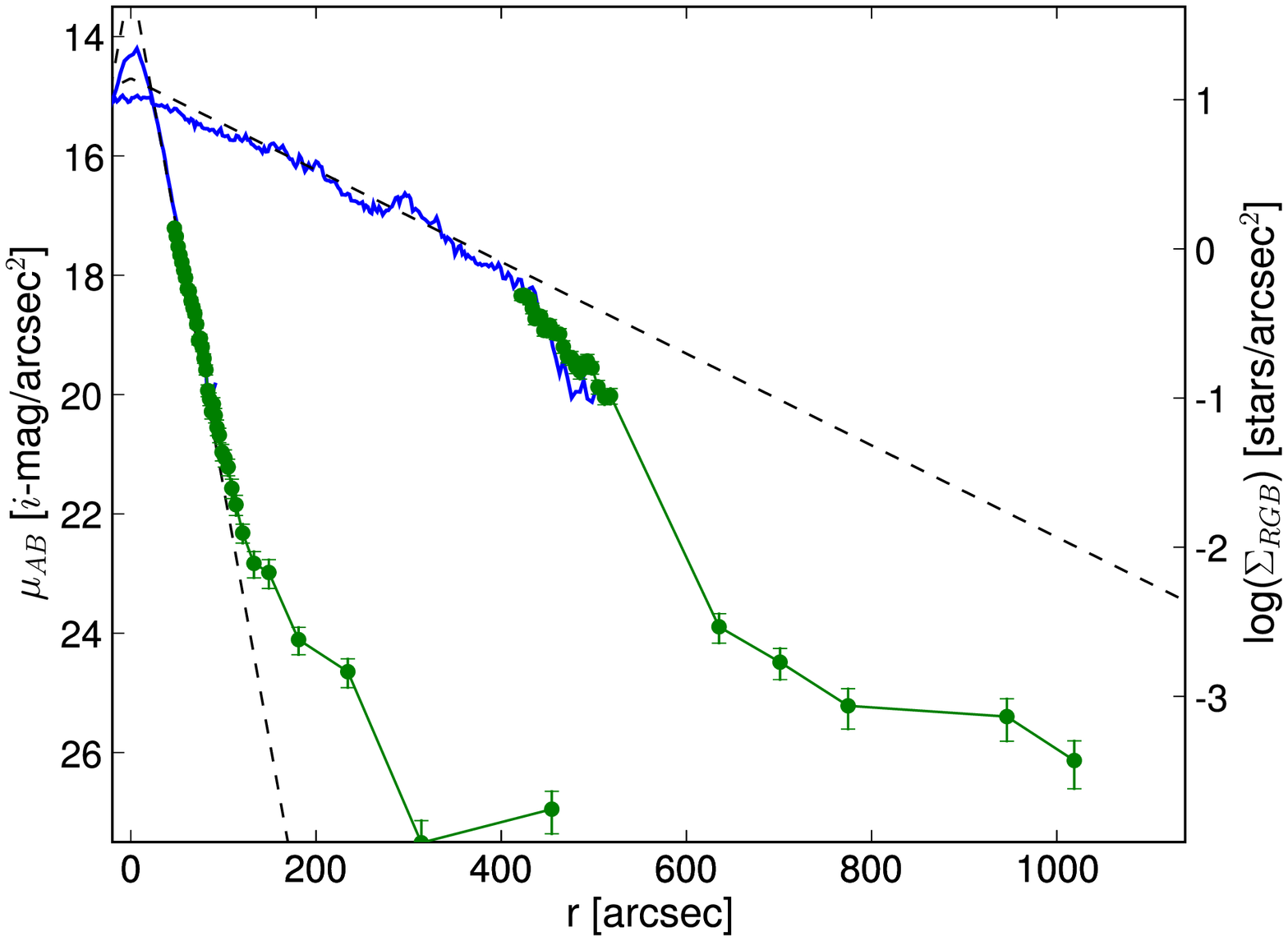}
\vspace{-5mm}
\caption{NGC\,4244 major and minor axis SDSS $i_{\rm AB}$-band
  luminosity profile (blue solid line, left axis, add about 6.5 to get
  $V$-band) and background subtracted star counts (green points, right
  axis). Dashed lines show exponential disk fits to the inner
  region. We detect a clear minor axis extended component (Seth et
  al.~2007) and a strong truncation in RGB star counts on the major
  axis. At the distance of NGC4244, 100 arcsec equals about 1.8\,kpc.
\label{ngc4244}
}
\end{minipage}
\hfill
\begin{minipage}{0.47\linewidth}
\vspace{-4mm}
\includegraphics[width=\linewidth]{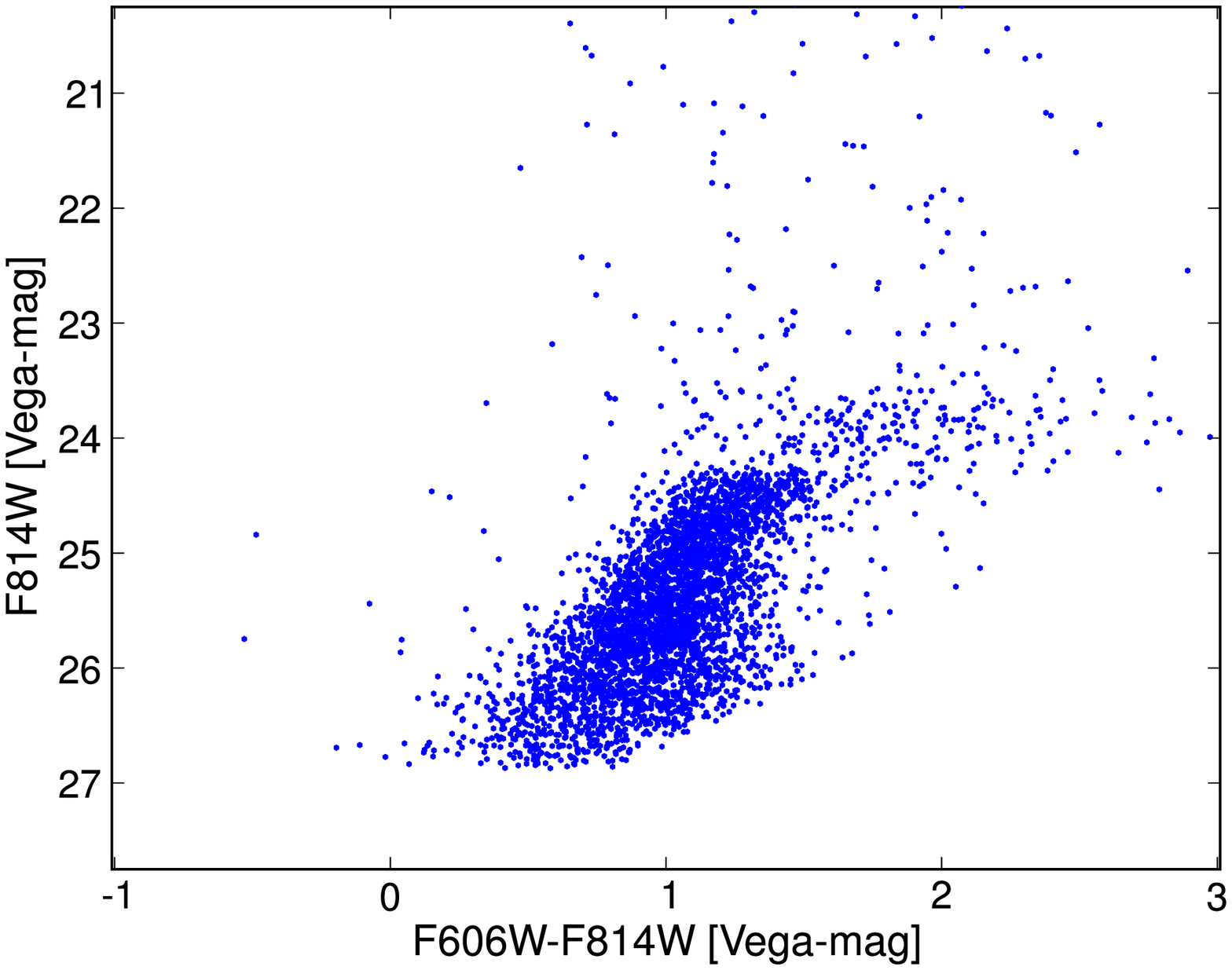}
\vspace{-5mm}
\caption{Color-magnitude diagram of the tidal stream around M83 at
  20\,kpc from its center. A very pronounced metal-poor RGB
  population is detected, with an AGB C-star population (F606W-F814W=1.3--2.5
  mag, F814W=24 mag), indicative of a 3--5\,Gyr old
  population. No main sequence or He burning stars are seen to the
  left of the RGB; this stream has been dead for at least 300\,Myr.
\label{m83}
}
\end{minipage}

\end{figure}

NGC\,4244 has a very small, very metal poor halo below $\mu_V$=30 mag
arcsec$^{-2}$ on the minor axis (Seth et al.\ 2007;
Fig.\ref{ngc4244}).  In contrast, most massive galaxies, like
NGC\,253, NGC\,891, and M94 have very extended halos; our outermost
fields at $\sim$30\,kpc still have $\mu_V$$\sim$28 mag
arcsec$^{-2}$. M81 has a projected $r^{-3.5}$ power-law minor axis
surface brightness profile, one of the steepest ever seen.
% (e.g., Zibetti et al.~2004 and reference therein). 
M83 seems to be the exception to the expectation that massive galaxies
have large halos; at 20\,kpc its CMD is already rather sparse.
The metallicities of the halos derived from the colors of RGB stars
are quite varied, although with more massive galaxies having higher
metallicity inner halos, on average.
% All galaxies show a trend of decreasing metallicity with radius. 
%Metallicities might be related to bulge size; bulge dominated
%M94 has at 15\,kpc a stellar population that is dominated by Solar
%metallity stars, with very few low metallicity stars.

The RGB stars in the thick disk of NGC\,891 and NGC\,4244 show a
truncation at the same radius as the total light distribution
(Fig.\,\ref{ngc4244}), suggesting that either truncations are old, or
that the old thick disk is affected by similar dynamical effects as
the thin disk (bars, spiral arms, disk heating and stripping by dark
matter subhalos).

We detect a stream in M83 with a maximum surface brightness of 26.5
$R$-mag arcsec$^{-2}$ and FWHM of $\sim$3\,kpc that has no detectable main
sequence nor He burning stars and therefore has had no star formation
in the past 300\,Myr.  However, we find a significant population of AGB
C-stars, indicating it had a burst of star formation about 3--5\,Gyr
ago (Fig.\,\ref{m83}).

\begin{acknowledgments}
Support for Program numbers GO-10523 and GO-10889 was provided by NASA through 
a grant from the Space Telescope Science Institute, which is operated by the 
Association of Universities for Research in Astronomy, Incorporated, under 
NASA contract NAS5-26555.

\end{acknowledgments}

\end{document}